# Tridiagonal Representation Approach in Quantum Mechanics

A. D. Alhaidari[a] and H. Bahlouli[b]

[a] *Saudi Center for Theoretical Physics, P.O. Box 32741, Jeddah 21438, Saudi Arabia*
[b] *Physics Department, King Fahd University of Petroleum & Minerals, Dhahran 31261, Saudi Arabia*

**Abstract**: We present an algebraic approach for finding exact solutions of the wave equation. The approach, which is referred to as the Tridiagonal Representation Approach (TRA), is inspired by the J-matrix method and based on the theory of orthogonal polynomials. The class of exactly solvable problems in this approach is larger than the conventional class. All properties of the physical system (energy spectrum of the bound states, phase shift of the scattering states, energy density of states, etc.) are obtained in this approach directly and simply from the properties of the associated orthogonal polynomials.

**PACS**: 03.65.Pm, 03.65.Ge, 03.65.Fd, 03.65.Ca

**Keywords**: tridiagonal representation, orthogonal polynomials, recursion relation, asymptotics, energy spectrum, phase shift.

## I. INTRODUCTION

A full understanding of the features and behavior of a physical submicroscopic system requires knowledge of the exact solution of the corresponding wave equation. This is especially evident when the system is in a critical state such as at phase transitions, singular limits, strong coupling, or when the physical parameters assume critical values, etc. In such circumstances, one may not be able to achieve full or correct understanding of the system by using numerical solutions. Now, since most of the quantum mechanical systems and physical processes are modeled by potential functions, then exact solutions of the wave equation with as many potential functions as possible remained as one of the prime challenges since the early inception of quantum mechanics. Additionally, exact solutions could also be used to test the accuracy and convergence of computational routines that were developed to obtain numerical solutions for complicated physical systems. Nonetheless, the class of exactly solvable potentials for a given wave equation (e.g., Schrödinger equation, Dirac equation, etc.) is indeed very limited. In the physics and mathematics literature, the potential functions in this class are well known and their exact solutions are well established. For example, the class of exactly solvable potentials in the nonrelativistic Schrödinger equation includes the Coulomb, harmonic oscillator, Morse, Pöschl-Teller, Scarf, Eckart, etc.

Several methods for obtaining exact solutions of the wave equation were developed over time. Most agree very well in their exact results but differ in accuracy and convergence when used to obtain approximations or numerical solutions of problems that are not exactly solvable. Among these methods, we mention supersymmetry [1], shape invariance [2], group theory [3-5], factorization [6,7], asymptotic iteration [8,9], point canonical transformation [10], path integral [11], Nikiforov-Uvarov [12], etc. Some of these are equivalent to each other but most (if not all) address the same class of exactly solvable potentials. Moreover, some are analytic methods while others are algebraic. The latter are usually preferred in numerical calculations especially if the former causes instabilities. The Tridiagonal Representation Approach (TRA) was introduced in 2005 to obtain exact solutions of the wave equation [13]. It is an algebraic method based on the theory of orthogonal polynomials and was inspired by the J-matrix method [14]. It has been applied to relativistic as well as nonrelativistic problems and in several spatial dimensions with separable potentials. The class of exactly solvable potentials in the TRA turns out to be larger than the conventional class. It includes new potential functions and generalizations of known ones. These potentials correspond to orthogonal polynomials that were not treated in the mathematics literature before. As an example of such potentials, we mention the infinite square potential well with sinusoidal rather than flat bottom. Another example, is a generalization of the hyperbolic Pöschl-Teller potential obtained by adding a term of the form $\tanh^2(x)/\cosh^2(x)$. A partial list of these new or generalized potentials is given in Table 1 of Ref. [15]. Some of these new solutions lead to interesting applications in atomic, molecular and nuclear physics. As examples, we mention the anion problem where an electron becomes bound to a neutral molecule with an electric dipole moment [16,17], the binding of a charged particle to an electric quadrupole in two dimensions [18], energy density bands engineering [19], electric dipole and quadrupole contributions to valence electron binding in a charge-screening environment [20], etc. In applied mathematics, the TRA was also used to obtain series solutions of new types of ordinary differential equations of the second order with three and four singular points [21-23]. In the following section, we start by introducing and formulating the TRA and explain its two working modes.

## II. FORMULATION OF THE TRA

The general spectrum of a quantum mechanical system is a mix of continuous and discrete energy states. The complete space-time wavefunction that represents such a system could be written in terms of its continuous and discrete Fourier components, in a standard notation, as follows

$$\Psi(x,t) = \int_\Omega e^{-iEt}\psi(E,x)dE + \sum_k e^{-iE_k t}\psi_k(x), \qquad (1)$$

where $\Omega$ is the set of continuous scattering energy interval(s) and $\{E_k\}$ is the countably infinite or finite set of discrete bound state energies. The continuous and discrete wavefunction components $\psi(E,x)$ and $\psi_k(x)$ could be considered as elements in an infinite dimensional vector space spanned by the basis vectors $\{\phi_n(x)\}$. In physics, we are accustomed to writing vector quantities (e.g., force, velocity, electric field, etc.) in terms of their components (i.e., projections on some conveniently chosen basis unit vectors in the space). For example, the force $\vec{F}$ is written in three dimensional space with Cartesian coordinates as $\vec{F} = f_x\hat{x} + f_y\hat{y} + f_z\hat{z}$, where $\{f_x, f_y, f_z\}$ are the projections of the force along the unit vectors $\{\hat{x}, \hat{y}, \hat{z}\}$. These projections contain all information about the physical quantity whereas the unit vectors (basis) are dummy, but must form a complete set to allow for a faithful representation of the vector quantity. In analogy, we write the wavefunction components $\psi(E,x)$ and $\psi_k(x)$ as a series expansion in terms of the complete set $\{\phi_n(x)\}$ of local unit vectors (square integrable basis functions in configuration space). That is, we write

$$|\psi(E,x)\rangle = \sum_n f_n(E)|\phi_n(x)\rangle, \qquad (2a)$$

$$|\psi_k(x)\rangle = \sum_n g_n(k)|\phi_n(x)\rangle, \qquad (2b)$$

where $f_n(E)$ and $g_n(k)$ are the projections of the corresponding wavefunction component on the basis unit vector $\phi_n(x)$. That is, $f_n(E) = \langle \bar{\phi}_n(x)|\psi(E,x)\rangle$ and $g_n(k) = \langle \bar{\phi}_n(x)|\psi_k(x)\rangle$, where $\bar{\phi}_n(x)$ is the conjugate of $\phi_n(x)$ defined so that $\langle \phi_n(x)|\bar{\phi}_m(x)\rangle = \langle \bar{\phi}_n(x)|\phi_m(x)\rangle = \delta_{nm}$. All physical information about the system (structure and dynamics) are contained in these projections. An alternative approach that leads to the same expansion of the wavefunction (2) is given in Appendix A. It also shows how to choose the *proper* basis set $\{\phi_n(x)\}$.

Faithfulness of the representation of the system by the total wavefunction $\Psi(x,t)$ means that its space correlation over all times is as follows

$$\int \Psi(x,t)\bar{\Psi}(y,t)dt = \delta(x-y), \qquad (3)$$

where $\bar{\Psi}(y,t)$ is the conjugate wavefunction, which is obtained from (1) and (2) by $i \to -i$ and $\phi_n(x) \to \bar{\phi}_n(x)$. Using the fact that the discrete energy spectrum and the continuous energy interval(s) are distinct (do not overlap), we conclude that $\int e^{\pm i(E_k - E)t} dt = 0$. Thus, the overlap integral between the discrete and continuous energy components in Eq. (3) vanishes and we obtain

$$\sum_{n,m} \phi_n(x)\bar{\phi}_m(y)\left[\int_\Omega f_0^2(E) P_n(E) P_m(E) dE + \sum_k g_0^2(k) Q_n(k) Q_m(k)\right] = \delta(x-y). \qquad (4)$$

where we have written, in anticipation of future convenience, $f_n(E) = f_0(E) P_n(E)$ and $g_n(k) = g_0(k) Q_n(k)$ making $P_0(E) = Q_0(k) = 1$. Consequently, the completeness of the basis, which reads $\sum_n \phi_n(x)\bar{\phi}_n(y) = \delta(x-y)$ dictates that

$$\int_\Omega \rho(E) P_n(E) P_m(E) dE + \sum_k \omega(k) Q_n(k) Q_m(k) = \delta_{nm}, \qquad (5)$$

where $\rho(E) = f_0^2(E)$ and $\omega(k) = g_0^2(k)$ are positive definite entire functions. This represents the completeness formula for the set $\{P_n(E), Q_n(k)\}$. Therefore, we can write the series expansion (2) as follows

$$|\psi(E,x)\rangle = \sqrt{\rho(E)} \sum_n P_n(E)|\phi_n(x)\rangle, \qquad (6a)$$

$$|\psi_k(x)\rangle = \sqrt{\omega(k)} \sum_n Q_n(k)|\phi_n(x)\rangle. \qquad (6b)$$

All physical information (both structural and dynamical) about the system are obtained from the properties of the set $\{P_n(E), Q_n(k)\}$. The initial values $P_0(E) = Q_0(k) = 1$ and orthogonality relation (5) suggest that this set represents a complete set of orthogonal polynomials with continuous and discrete elements $\{P_n(E)\}$ and $\{Q_n(k)\}$ whose corresponding weight functions are $\rho(E)$ and $\omega(k)$, respectively. It turns out that the type of orthogonal polynomials representing the physical system depends on the structure of its energy spectrum: purely continuous, purely discrete or a mix of both and on whether the discrete energy spectrum is infinite or finite. Table 1 shows this relationship with examples. These premonitions will be confirmed by the ensuing development below.

To simplify the presentation of the TRA to a general audience of quantum mechanics including undergraduate students, we limit the discussion in the following two sections (Secs. III and IV) to physical systems with either pure discrete energy bound states or pure continuous energy scattering states. For the more advanced readers, however, we give in section V an example of a system with mixed energy spectrum. Now, let the Hamiltonian operator of the system in configuration space be $H(x) = T(x) + V(x)$, where $T$ is the kinetic energy operator and $V$ is the potential function. The wave equation is $i\hbar \frac{\partial}{\partial t} \Psi(x,t) = H(x)\Psi(x,t)$ and thus the continuous or discrete Fourier components of the wave equation read as follows

$$H(x)|\psi(x)\rangle = E|\psi(x)\rangle, \tag{7}$$

where $E$ stands for either continuous energies corresponding to the scattering states $\psi(E,x)$ or discrete energies $\{E_k\}$ corresponding to the bound states $\{\psi_k(x)\}$. Inserting the series expansion (6) in the wave equation (7) and projecting from left by $\langle \phi_n(x)|$ result in the following equivalent matrix equation for the orthogonal energy polynomials

$$\sum_m \left[ \langle \phi_n | H | \phi_m \rangle - E \langle \phi_n | \phi_m \rangle \right] P_m^\mu(E) = 0, \tag{8}$$

where $\mu$ stands for a set of physical parameters contained in the Hamiltonian. This equation could be rewritten as $\sum_m J_{n,m}(E) P_m^\mu(E) = 0$, where $J_{n,m}(E)$ is the matrix representation of the wave operator $J(x) = H(x) - E$ in the $L^2$ basis $\{\phi_n(x)\}$. Now, we come to the most crucial step in the TRA and that is to impose the condition on the basis that it must produce a tridiagonal matrix representation for the wave operator $J(E)$. Moreover, since the Hamiltonian is Hermitian then the matrix $J(E)$ should also be symmetric. That is, the matrix wave equation (8) must read

$$J_{n,n-1}(E) P_{n-1}^\mu(E) + J_{n,n}(E) P_n^\mu(E) + J_{n,n+1}(E) P_{n+1}^\mu(E) = 0. \tag{9}$$

This could always be reparametrized and rewritten as follows:

$$z c_n P_n^\mu(z) = a_n P_n^\mu(z) + b_{n-1} P_{n-1}^\mu(z) + b_n P_{n+1}^\mu(z), \tag{10}$$

where $n = 1, 2, 3, ...$ and $z$ is some proper function of the energy $E$ and the set of physical parameters $\mu$. If the basis elements are orthonormal (i.e., $\langle \phi_n | \phi_m \rangle = \delta_{nm}$) then $z = E$ and $c_n = 1$ but this is not always the case. The coefficients $\{a_n, b_n, c_n\}$ depend on $\mu$ and $n$ but are independent of $z$ and such that $b_n^2 > 0$ for all $n$. Equation (10) is, in fact, a recursion relation that

–4–

makes $P_n^\mu(z)$ a polynomial of degree $n$ in $z$ and determines all of them explicitly to any degree starting with the initial values $P_0^\mu(z) = 1$ and $P_1^\mu(z) = (zc_0 - a_0)/b_0$. Therefore, using the theory of orthogonal polynomials we can assert the following properties of the set $\{P_n(E), Q_n(k)\}$ introduced above in Eq. (4) and Eq. (5)

1. $\{P_n(E)\}$ constitutes a set of orthogonal polynomials with continuous spectrum whose discrete version is $\{Q_n(k)\}$.
2. Equation (5) is the generalized orthogonality relation with $\rho(E)$ and $\omega(k)$ being the continuous and discrete weight functions associated with $\{P_n(E)\}$ and $\{Q_n(k)\}$, respectively.

An example of orthogonal polynomial with purely continuous spectrum is the Meixner-Pollaczek polynomial whose discrete version is the Meixner polynomial (or Krawtchouk polynomial) with infinite (or finite) discrete spectrum [24]. Moreover, an orthogonal polynomial with a mix of continuous and discrete spectra whose generalized orthogonality relation has the same structure as Eq. (5) is the Wilson polynomial [24]. It is evident that if the system represented by the complete wavefunction $\Psi(x,t)$ possesses only continuous energy scattering states, then the sum part of the orthogonality relation (5) disappears. However, if the system consists only of discrete energy bound states then the integral part of the orthogonality relation (5) disappears. These findings are supported by studies that established the connection between scattering theory and orthogonal polynomials [25-27] and by the recent extensive work on the formulation of quantum mechanics based on the theory of orthogonal polynomials (see, for example, Ref. [15,28-30]).

The well-established connection between scattering and orthogonal polynomials dictates that out of all orthogonal polynomials that satisfy the above requirements, the physically relevant ones that enter in the expansion of the wavefunction (6a) are those with the following asymptotic ($n \to \infty$) behavior

$$P_n^\mu(z) \approx n^{-\tau} A^\mu(z) \left\{ \cos\left[ n^\xi \theta(z) + \varphi(z) \log n + \delta^\mu(z) \right] + O(n^{-1}) \right\}, \qquad (11)$$

where $\tau$ and $\xi$ are real positive constants that depend on the particular energy polynomial. In most cases, either $\varphi(z) = 0$ or $\theta(z) = 0$. The studies in [15,25-30] show that $A^\mu(z)$ is the scattering amplitude and $\delta^\mu(z)$ is the phase shift. Bound states, if they exist, occur at discrete real energies $\{E(z_k)\}$ that make the scattering amplitude vanish, $A^\mu(z_k) = 0$. The number of these bound states is either finite or infinite.

The main task in the TRA is to find the energy polynomials with continuous and/or discrete spectrum associated with the given physical system because (as noted above and restated here for emphasis) all physical information about the system are obtained from the properties of these polynomials. There are two working modes of the TRA. In the first one, a potential function that models the given system is provided and the TRA is used to obtain the corresponding exact solution (i.e., the orthogonal energy polynomial). In the second mode, a general physical configuration is given and the TRA is used to obtain the class of all exactly solvable potential functions that are compatible with the configuration. For the first mode, we search for a complete square integrable basis set $\{\phi_n\}$ that satisfies the boundary conditions and supports a symmetric tridiagonal matrix representation for the wave operator with the given potential. There is no prior guarantee that the task will be successful, but if it is, then we derive or identify the orthogonal polynomial that satisfies the corresponding three-term recursion relation. On the other hand, for the second mode we start by choosing a proper general basis that is compatible with the given physical configuration then compute $\langle\phi_n|\phi_m\rangle$ and the matrix elements of the kinetic energy operator $\langle\phi_n|T|\phi_m\rangle$. Afterwards, we find all possible potential functions, $V(x)$, that leave the matrix representation of the wave operator, $\langle\phi_n|T|\phi_m\rangle + \langle\phi_n|V|\phi_m\rangle - E\langle\phi_n|\phi_m\rangle$, tridiagonal. Finally, for each potential function in the class we derive or identify the orthogonal polynomial that satisfies the resulting three-term recursion relation. Subsequently, all properties of the physical system corresponding to a given potential function in the class are derived from the properties of the associated orthogonal polynomial. For example, we can use the asymptotics formula (11) to obtain the scattering amplitude and phase shift for the continuum scattering states as well as the energy spectrum for the discrete bound states. For clarity and simplicity of the presentation, we leave out details of the calculation but interested readers can find such details in [21,31].

We should note that all *new* and generalized potential functions that are solved exactly using the TRA but have no exact solution using the conventional methods correspond to orthogonal polynomials that are either new or modified/generalized versions of known polynomials. Most of these polynomials were not studied in the mathematics literature and some still constitute an open problem in orthogonal polynomials [32,33]. In such cases and due to the absence of the analytic properties of these polynomials (such as the weight function, asymptotic formula, etc.), we are forced to resort to numerical means to obtain the physical properties of the system.

In sections III, we present an example of how to use the TRA in the first mode of its application and obtain an exact solution for a conventional potential function (the Coulomb). For simplicity, we consider only scattering where the energy spectrum is purely continuous. In section IV, we obtain a solution of a non-conventional potential with purely discrete energy spectrum: the infinite square well with sinusoidal rather than flat bottom. In section V and for the more advanced readers, we present a system with mixed spectrum consisting of continuous scattering states and discrete bound states. In Appendix B, we present an example of how to use the TRA in the second mode of its application and obtain the class of all exactly solvable potentials corresponding to a given physical configuration. Some of these systems are exactly solvable only by using the TRA.

## III. TRA SOLUTION FOR A CONVENSIONAL POTENTIAL FUNCTION

To ease the way into the TRA, we start by showing how the approach could be used in its first mode of application to obtain the exact solution of a well-known problem; the continuous energy scattering states of the Coulomb problem with $V(r) = Z/r$, where the electric charge number $Z$ is positive. In the atomic units $\hbar = m = \frac{e^2}{4\pi\varepsilon_0} = 1$, the radial Schrödinger wave equation for this problem reads as follows

$$\left[-\frac{1}{2}\frac{d^2}{dr^2} + \frac{\ell(\ell+1)}{2r^2} + \frac{Z}{r} - E\right]\psi(E,r) = 0, \qquad (12)$$

where $\ell$ is the angular momentum quantum number. This equation has a regular singularity at $r = 0$ and an irregular (essential) singularity at infinity. Using this fact and that $r \geq 0$, the analysis in Appendix A shows that a proper choice of a complete set of square integrable basis that satisfy the boundary conditions is

$$\phi_n(r) = A_n y^\alpha e^{-y/2} L_n^\nu(y), \tag{13}$$

where $y = \lambda r$, $L_n^\nu(y)$ is the associated Laguerre polynomial, $A_n$ is a normalization constant, $\lambda$ is a positive length scale parameter and $(\alpha, \nu)$ are real dimensionless parameters such that $\alpha > 0$ and $\nu > -1$. Using the differential equation and differential property of the Laguerre polynomial, we obtain the following action of the wave operator on the basis

$$J|\phi_n\rangle = \frac{\lambda^2}{2}\left[\frac{n}{y}\left(1 + \frac{\nu + 1 - 2\alpha}{y}\right) + \frac{\ell(\ell+1) - \alpha(\alpha-1)}{y^2} + \frac{\alpha + 2\gamma}{y} - \varepsilon - \frac{1}{4}\right]|\phi_n\rangle$$
$$+ \frac{\lambda^2}{2}\left[(2\alpha - \nu - 1)\frac{n+\nu}{y^2}\right]\frac{A_n}{A_{n-1}}|\phi_{n-1}\rangle \tag{14}$$

where $\gamma = Z/\lambda$ and $\varepsilon = (\kappa/\lambda)^2$ with $\kappa$ being the linear momentum $\kappa^2 = 2E$. The recursion relation of the Laguerre polynomial shows that the matrix representation of the wave operator $\langle \phi_m | J | \phi_n \rangle$ is tridiagonal if the terms with $y^{-2}$ in (14) vanish. Thus, we require that $2\alpha = \nu + 1$ and $\nu^2 = (2\ell+1)^2$ giving $\nu = 2\ell+1$ and $\alpha = \ell+1$. Consequently, we obtain

$$\frac{2}{\lambda^2}\langle \phi_m | J | \phi_n \rangle = (n + \ell + 1 + 2\gamma)\langle \phi_m | y^{-1} | \phi_n \rangle - \left(\varepsilon + \tfrac{1}{4}\right)\langle \phi_m | \phi_n \rangle. \tag{15}$$

Using the recursion relation of the Laguerre polynomial and its orthogonality along with $A_n = \sqrt{\Gamma(n+1)/\Gamma(n+2\ell+2)}$, we obtain

$$\frac{2}{\lambda^2} J_{n,m} = (n + \ell + 1 + 2\gamma)\delta_{n,m}$$
$$-\left(\varepsilon + \tfrac{1}{4}\right)\left[2(n+\ell+1)\delta_{n,m} - \sqrt{n(n+2\ell+1)}\delta_{n,m+1} - \sqrt{(n+1)(n+2\ell+2)}\delta_{n,m-1}\right] \tag{16}$$

Therefore, the symmetric three-term recursion relation (9) becomes

$$\frac{-2\gamma}{\varepsilon + 1/4}P_n^\mu(E) = -2(n+\ell+1)\frac{\varepsilon - 1/4}{\varepsilon + 1/4}P_n^\mu(E)$$
$$\sqrt{n(n+2\ell+1)}P_{n-1}^\mu(E) + \sqrt{(n+1)(n+2\ell+2)}P_{n-1}^\mu(E) \tag{17}$$

Comparing this to the three-term recursion relation of the normalized version of the Meixner-Pollaczek polynomial $\mathcal{P}_n^\sigma(z,\theta)$, which is given by Eq. (D2) of Appendix D in Ref. [31], we can write $P_n^\mu(E) = \mathcal{P}_n^{\ell+1}(z,\theta)$, where

$$\cos\theta = \frac{\varepsilon - 1/4}{\varepsilon + 1/4} = \frac{(2\kappa/\lambda)^2 - 1}{(2\kappa/\lambda)^2 + 1}, \qquad z = -\gamma/\sqrt{\varepsilon} = -Z/\kappa. \tag{18}$$

The normalized version of the Meixner-Pollaczek polynomial is written in terms of the hypergeometric function as given by Eq. (D1) in Appendix D of Ref. [31]. Thus, the scattering wave function (6a) is written as

$$\psi(E,r) = (\lambda r)^{\ell+1} e^{-\lambda r/2} \sqrt{\rho^{\ell+1}(-Z/\kappa,\theta)} \sum_{n=0}^{\infty} A_n \mathcal{P}_n^{\ell+1}(-Z/\kappa,\theta) L_n^{2\ell+1}(\lambda r). \tag{19}$$

where $\rho^\sigma(z,\theta)$ is the positive definite weight function associated with the Meixner-Pollaczek polynomial $\mathcal{P}_n^\sigma(z,\theta)$ and given by Eq. (D3) in Appendix D of Ref. [31]. In practice, the sum in (19) converges very quickly with the desired accuracy for the first few terms. The asymptotics of $\mathcal{P}_n^\sigma(z,\theta)$ is given by Eq. (D4) in Appendix D of Ref. [31]. It has exactly the same form as Eq. (11) from which the scattering phase shift (modulo $\pi/2$) is obtained directly and effortlessly as

$$\delta(E) = \arg\left[\Gamma(\ell + 1 - iZ/\kappa)\right], \tag{20}$$

which is the well-known scattering phase shift for the Coulomb problem.

### IV. TRA SOLUTION FOR A NON-CONVENSIONAL POTENTIAL FUNCTION

Now, we give an example that illustrates how the TRA could be used to solve a problem that is not exactly solvable using the conventional methods. One of the first problems that an undergraduate student is quantum mechanics is asked to solve is the infinite square potential well. Here, we add a twist on this problem and look for its solution when the bottom of the well is not flat but rather sinusoidal. Specifically, we want to solve the problem with the following potential function

$$V(x) = \begin{cases} V_0 \sin(\pi x/L) & , -\frac{L}{2} \le x \le +\frac{L}{2} \\ \infty & , \text{otherwise} \end{cases} \tag{21}$$

Since the system is totally confined in space, then it should be obvious that it has an infinite number of discrete energy bound states $\{\psi_k(x)\}$ and no continuous energy states. With modest efforts, it is not difficult to convince oneself that an exact solution for this potential is not achievable using any of the traditional methods (e.g., supersymmetry, factorization, asymptotic iteration, etc.). Now, we employ the TRA and start by writing the corresponding Schrödinger equation in terms of the dimensionless coordinate $y = \sin(\lambda x)$ with $\lambda = \pi/L$ where we obtain

$$-\frac{\lambda^2}{2}\left[(1-y^2)\frac{d^2}{dy^2} - y\frac{d}{dy} - \gamma y + \varepsilon\right]\psi_k(x) = 0, \tag{22}$$

with $\gamma = 2V_0/\lambda^2$, $\varepsilon = 2E/\lambda^2$ and $-1 \leq y \leq +1$. This equation has two regular singularities at $y = \pm 1$. Using these facts, the analysis in Appendix A shows that a proper choice of a complete set of square integrable basis that satisfies the boundary conditions is

$$\phi_n(x) = \sqrt{\tfrac{2}{\pi}}(1-y^2)^\alpha U_n(y), \tag{23}$$

where $U_n(y)$ is the Chebyshev polynomial of the second kind and $\alpha$ is a real positive dimensionless parameter. Using the second order differential equation for $U_n(y)$, we obtain the following action of the wave operator on this basis

$$J|\phi_n\rangle = -\frac{\lambda^2}{2}(1-y^2)^\alpha \left[2(1-2\alpha)y\frac{d}{dy} + \frac{2\alpha(2\alpha-1)y^2}{1-y^2} - \gamma y - n(n+2) - 2\alpha + \varepsilon\right]U_n(y) \tag{24}$$

This action is tridiagonal if the right-hand side of this equation contains terms proportional only to $\phi_n$ and $\phi_{n\pm 1}$ with constant coefficients. The recursion relation of the Chebyshev polynomial shows that this is possible only if the terms inside the square brackets in (24) are linear in $y$. Thus, we require that $\alpha = \tfrac{1}{2}$ resulting in the following matrix representation of the wave operator

$$\frac{2}{\lambda^2}\langle\phi_m|J|\phi_n\rangle = \gamma\langle\phi_m|y|\phi_n\rangle + \left[(n+1)^2 - \varepsilon\right]\langle\phi_m|\phi_n\rangle. \tag{25}$$

Note that $\langle\phi_m|F(y)|\phi_n\rangle$ for a given function $F(y)$ is calculated as the following integral

$$\langle\phi_m|F(y)|\phi_n\rangle = \lambda\int_{-L/2}^{+L/2}\phi_m(x)F(y)\phi_n(x)dx = \tfrac{2}{\pi}\int_{-1}^{+1}(1-y^2)^{2\alpha}U_m(y)F(y)U_n(y)\frac{dy}{\sqrt{1-y^2}}. \tag{26}$$

Using the recursion relation of $U_n(y)$ and its orthogonality, we obtain the following symmetric three-term recursion relation (9)

$$\varepsilon P_n^\mu(E) = (n+1)^2 P_n^\mu(E) + \tfrac{1}{2}\gamma\left[P_{n-1}^\mu(E) + P_{n+1}^\mu(E)\right]. \tag{27}$$

It should be obvious that the infinite square potential well with flat bottom (where $\gamma = 0$) is a special case of (27) that results in a diagonal representation giving the well-known energy spectrum directly as $\varepsilon_n = (n+1)^2$. On the other hand, for $\gamma \neq 0$ we compare this recursion relation to that of the newly introduced orthogonal polynomial $H_n^{(\mu,\nu)}(z,\sigma)$, which is given as Eq. (E1) of Appendix E in Ref. [31]. We then conclude that $P_n^\mu(E) = H_n^{(\frac{1}{2},\frac{1}{2})}(\varepsilon,\gamma)$ and that the bound state wavefunction (6b) is written as

$$\psi_k(x) = \sqrt{\tfrac{2}{\pi}}\cos(\lambda x)\sqrt{\omega^{(\frac{1}{2},\frac{1}{2})}(\varepsilon_k,\gamma)}\sum_{n=0}^{\infty} H_n^{(\frac{1}{2},\frac{1}{2})}(\varepsilon_k,\gamma)U_n(\sin(\lambda x)), \tag{28}$$

–9–

where $\omega^{(\mu,\nu)}(z,\sigma)$ is the positive definite weight function associated with the new polynomial $H_n^{(\mu,\nu)}(z,\sigma)$. Unfortunately, the analytic properties of this polynomial (like its weight function, orthogonality, asymptotics, etc.) are not yet known. This is still an open problem in orthogonal polynomials [32,33] despite the fact that this polynomial could be written explicitly to any desired degree (albeit not in closed form) using its recursion relation and initial value $H_0^{(\mu,\nu)}(z,\sigma)=1$. Had the analytic properties of this polynomial been known, we would have easily and directly obtained the physical properties of the present system (e.g., the energy spectrum). Consequently, we are forced to resort to numerical means. In Table 2, we give the lowest part of the energy spectrum (in units of $\frac{1}{2}\lambda^2$) calculated for a chosen set of values of the potential parameter $\gamma$. Note that for $\gamma = 0$ we obtain the well-known result and for $\gamma \neq 0$ the higher excited states do not feel the sinusoidal structure at the bottom of the well. In Figure 1, we plot the lowest un-normalized bound-state wavefunctions corresponding to the column $\gamma = 5$ of Table 2. These were calculated using Eq. (28) where the sum converges quickly for the first 7-8 terms.

In Appendix B, we use the second mode of application of the TRA to find the class of all potential functions associated with the physical configuration of the infinite square well and show that the potential well with sinusoidal bottom treated above is only a special case of a larger class (the generalized trigonometric Scarf potential).

## V. TRA SOLUTION FOR A SYSTEM WITH MIXED SPECTRUM

In this section, we use the TRA to obtain the exact solution of the Schrödinger equation for a system with mixed spectrum consisting of discrete energy bound states and continuous energy scattering states. We give the condition(s) for the existence of bound states and use the properties of the associated orthogonal energy polynomials to write down analytic expressions for the bound states energy spectrum and the scattering states phase shift. As an illustration, we choose the one-dimensional Morse potential $V(x) = V_2 e^{2\lambda x} + V_1 e^{\lambda x}$, where $-\infty < x < +\infty$ and $\{\lambda, V_2, V_1\}$ are real parameters with $\lambda$ being a positive length scale. In the atomic units $\hbar = m = 1$, the corresponding Schrödinger equation in terms of the variable $y = e^{\lambda x}$ reads as follows

$$-\frac{\lambda^2}{2}\left[y^2 \frac{d^2}{dy^2} + y\frac{d}{dy} - U_2 y^2 - U_1 y + \varepsilon\right]\psi(E,x) = 0, \qquad (29)$$

where $U_i = 2V_i/\lambda^2$. This equation has a regular singularity at $y = 0$ and an irregular singularity at $y \to +\infty$. Using this fact and that $y \geq 0$, the analysis in Appendix A shows that a proper choice of a complete set of square integrable basis that satisfy the boundary conditions is

$$\phi_n(x) = A_n y^\alpha e^{-y/2} L_n^\nu(y). \qquad (30)$$

The real parameters $(\alpha, \nu)$ are dimensionless such that $\alpha > 0$ and $\nu > -1$. Using the differential equation and differential property of the Laguerre polynomial, we obtain the following action of the wave operator on the basis

$$J|\phi_n\rangle = -\frac{\lambda^2}{2}\left[y^2\left(\tfrac{1}{4}-U_2\right) - y\left(n+\alpha+\tfrac{1}{2}+U_1\right) + n(2\alpha-\nu) + \alpha^2 + \varepsilon\right]|\phi_n\rangle$$
$$-(2\alpha-\nu)(n+\nu)\frac{A_n}{A_{n-1}}|\phi_{n-1}\rangle \tag{31}$$

Using the recursion relation of the Laguerre polynomial, its orthogonality and that $\langle\phi_m|F(y)|\phi_n\rangle$, for a given function $F(y)$, is the following integral

$$\langle\phi_m|F(y)|\phi_n\rangle = \lambda\int_{-\infty}^{+\infty}\phi_m(x)F(y)\phi_n(x)dx = A_m A_n \int_0^\infty y^{2\alpha}e^{-y}L_m^\nu(y)F(y)L_n^\nu(y)\frac{dy}{y}, \tag{32}$$

where $A_n = \sqrt{\Gamma(n+1)/\Gamma(n+\nu+1)}$. We can show that Eq. (31) produces a symmetric tridiagonal matrix representation for $\langle\phi_m|J|\phi_n\rangle$ in either one of the following two cases

a) $2\alpha = \nu$, $\qquad \nu^2 = -4\varepsilon$. (33a)
b) $2\alpha = \nu+1$, $\qquad U_2 = \tfrac{1}{4}$. (33b)

The first case is valid for negative energies, which corresponds to a spectrum consisting only of bound states. Therefore, we dismiss this case and consider case (33b) corresponding to the following two-parameter 1D Morse potential

$$V(x) = \tfrac{\lambda^2}{8}e^{2\lambda x} + V_1 e^{\lambda x}. \tag{34}$$

Putting all the above together enables us to calculate the elements of the tridiagonal matrix representation of the wave operator, $\langle\phi_m|J|\phi_n\rangle$, which in turn results in the symmetric three-term recursion relation (9) that reads as follows

$$\varepsilon P_n^\mu(E) = \left[(2n+\nu+1)\left(n+\tfrac{\nu+1}{2}+U_1\right) - \tfrac{1}{4}(\nu^2-1)\right]P_n^\mu(E)$$
$$-\left(n+\tfrac{\nu}{2}+U_1\right)\sqrt{n(n+\nu)}P_{n-1}^\mu(E) - \left(n+1+\tfrac{\nu}{2}+U_1\right)\sqrt{(n+1)(n+\nu+1)}P_{n+1}^\mu(E) \tag{35}$$

Comparing this with the three-term recursion relation of the normalized version of the continuous dual Hahn polynomial $S_n^\tau(z^2;a,b)$, which is given by Eq. (B4) of Appendix B in Ref. [15], we can write $P_n^\mu(E) = S_n^{U_1+\frac{1}{2}}\left(\varepsilon;\tfrac{\nu+1}{2},\tfrac{\nu+1}{2}\right)$. The normalized version of the continuous dual Hahn polynomial is written in terms of the hypergeometric function $_3F_2$ as shown by Eq. (B1) in Appendix B of Ref. [15]. The properties of the polynomial $S_n^\tau(z^2;a,b)$ shows that if $\tau > 0$ then it has only a continuous spectrum with $z^2 > 0$. However, if $\tau < 0$ then the spectrum is a mix of continuous and discrete parts with the discrete part being of finite size $N$ where $N$ is the largest integer less than or equal to $-\tau$. Therefore, if $V_1 > -(\lambda/2)^2$ (i.e., $U_1 > -\tfrac{1}{2}$) then the potential cannot support bound states whereas if $V_1 < -(\lambda/2)^2$ (i.e., $U_1 < -\tfrac{1}{2}$) then it can and the system's total wavefunction has a continuous energy component and discrete energy components of a finite size as given by Eq. (1) with the sum being for $k = 0,1,...,N$ where $N$ is the largest integer less than or equal to $-U_1 - \frac{1}{2}$. The continuous and discrete energy components of the wavefunction are written as

$$\psi(E,x) = e^{\frac{\nu+1}{2}(\lambda x)} \exp\left(-\tfrac{1}{2}e^{\lambda x}\right)\sqrt{\rho^{U_1+\frac{1}{2}}\left(\varepsilon; \tfrac{\nu+1}{2}, \tfrac{\nu+1}{2}\right)} \sum_{n=0}^{\infty} A_n S_n^{U_1+\frac{1}{2}}\left(\varepsilon; \tfrac{\nu+1}{2}, \tfrac{\nu+1}{2}\right) L_n^{\nu}(e^{\lambda x}). \quad (36a)$$

$$\psi_k(x) = e^{\frac{\nu+1}{2}(\lambda x)} \exp\left(-\tfrac{1}{2}e^{\lambda x}\right)\sqrt{\omega^{U_1+\frac{1}{2}}\left(\varepsilon_k; \tfrac{\nu+1}{2}, \tfrac{\nu+1}{2}\right)} \sum_{n=0}^{\infty} A_n S_n^{U_1+\frac{1}{2}}\left(\varepsilon_k; \tfrac{\nu+1}{2}, \tfrac{\nu+1}{2}\right) L_n^{\nu}(e^{\lambda x}). \quad (36b)$$

where $\rho^{\tau}(z^2;a,b)$ is the continuous part of the weight function associated with $S_n^{\tau}(z^2;a,b)$ and given by Eq. (B2) in Appendix B of Ref. [15]. On the other hand, $\omega^{\tau}(z_k^2;a,b)$ is the discrete part of the weight function obtained from formula (B3) in Appendix B of Ref. [15], which has exactly the same form as Eq. (3) above. The asymptotic of $S_n^{\tau}(z^2;a,b)$ is given by Eq. (B5) in Appendix B of Ref. [15], which has the same form as Eq. (11) above with $\theta(z) = 0$. The scattering amplitude and scattering phase shift obtained from the asymptotics, which are shown as Eq. (B6) and Eq. (B7) in [15], give

$$E_k = -\frac{\lambda^2}{2}\left(k + \tfrac{1}{2} + \tfrac{2V_1}{\lambda^2}\right)^2. \quad (37)$$

$$\delta(E) = \arg\Gamma(2i\kappa/\lambda) - \arg\Gamma\left(\tfrac{1}{2} + \tfrac{2V_1}{\lambda^2} + i\kappa/\lambda\right) - 2\arg\Gamma\left(\tfrac{\nu+1}{2} + i\kappa/\lambda\right), \quad (38)$$

where $\kappa^2 = 2E$ and $k = 0,1,...,N$ with $N$ being the largest integer less than or equal to $-\left(\tfrac{1}{2} + 2V_1/\lambda^2\right)$. These are the well-known results associated with the 1D Morse potential (34) (see, for example, references [10,34,35]).

## VI. CONCLUSION

In summary, the Tridiagonal Representation Approach aims at utilizing the expansion of the wavefunction in a suitable square integrable basis set that transforms the usual Schrödinger equation into a three-term recursion relation for the expansion coefficients of the wavefunction. The recursion relation is then solved in terms of orthogonal polynomials, which are usually expressed as variants of the hypergeometric orthogonal polynomials (e.g., the Meixner-Pollaczek, continuous dual Hahn, Wilson, Racah, etc.). However, in some cases that correspond to new classes of solvable potentials, the orthogonal polynomials associated with the resulting recursion relation do not belong to the known polynomials. Consequently, their weight functions, generating functions, zeros, etc. are yet to be derived analytically. Nonetheless, they are completely defined by their three-term recursion relations and initial seed values. Consequently, in the absence of a closed-form solution of the recursion relation, we choose to use the term "algebraic" to describe the solutions obtained for this larger class of problems. However, these are *exact* solutions in the sense that all objects needed to calculate the sought after physical quantities in the problem (e.g., the energy spectrum, phase shift, wavefunction, density of states, etc.) are known and given to all orders. The accuracy in the values obtained for such quantities is limited only by the computing machine accuracy; no physical approximations are involved.

Using the TRA we have been able to obtain exact solutions of the Schrödinger equation by requiring a tridiagonal representation of the corresponding wave operator. As such, we were able to generate a larger class of regular solutions of the original Schrödinger equation. The advantage of the TRA is that the total energy spectrum of the problem both continuous and discrete are contained in a single associated energy polynomial. The approach is limited by the types of mapping of configuration space with coordinate $x$ into $y(x)$ that restricts the $y$-physical domain to be either infinite, semi-infinite or finite to enable us to use the appropriate classical orthogonal polynomials in the choice of bases as shown in Appendix A. Nonetheless, all known classes of exactly solvable potentials fall within the range of application of our approach.

To illustrate the TRA in its two modes of application, we considered scattering in the familiar Coulomb problem, which has a continuous energy spectrum where we derived the scattering phase shift. Then we presented an example of the infinite potential well with sinusoidal bottom, which is a confined system with only discrete energy spectrum and is exactly solvable only by using the TRA. Then in Appendix B, we showed that this potential well is only an element in a larger class of potentials represented by a three-parameter generalized trigonometric Scarf potential. Finally, using the unique features of orthogonal polynomials with mixed spectrum we presented, as an illustrative example, the two-parameter 1D Morse potential that supports both discrete energy bound states as well as continuous energy scattering states.

**ACKNOWLEDGEMENTS**

The support provided by the Saudi Center for Theoretical Physics (SCTP) is highly appreciated. We also acknowledge partial support by King Fahd University of Petroleum and Minerals (KFUPM).

**APPENDIX A: BASIS SET FOR THE EXPANSION OF THE WAVEFUNCTION**

In this Appendix, we give an alternative view for the origin of the wavefunction expansion given by Eq. (2) and show how to choose a proper square integrable basis $\{\phi_n(x)\}$ for a given problem. Performing a general coordinate transformation $x \to y(x)$ typically makes the wave equation a second order differential equation in $y$. For such an equation with regular singular points and due to Fuchs' theorem, we can use Frobenius method to write the following power series expansion for the solution

$$\psi(E,x) = (y-y_1)^{\alpha_1}(y-y_2)^{\alpha_2}....(y-y_r)^{\alpha_r}\sum_{n=0}^{\infty}c_n(y-\hat{y})^n, \tag{A1}$$

where $\{y_n\}$ are the singular points of the differential equation and $\hat{y}$ is one of these points or any other regular point. The exponents $\{\alpha_n\}$ are dimensionless parameter to be determined by the "indicial equation". The expansion coefficients $\{c_n\}$ depend on the parameters of the wave equation and the energy. In the expansion (A1), we assumed that the singular points are finite. However, if one of them is irregular (essential) at $+\infty$ then the factors multiplying the sum contains the term $e^{-\beta y}$ and if, on the other hand, two of them are irregular at $\pm\infty$ then the factors include $e^{-\beta y^2}$. Substituting (A1) into the differential wave equation will result in the indicial equation and a recurrence relation for the expansion coefficients. It is easy to see that we can

−13−

replace $(y-\hat{y})^n$ by a polynomial $p_n(y)$ of degree $n$ in $y$ with different expansion coefficients as follows

$$\psi(E,x) = R(y-\hat{y})\sum_{n=0}^{\infty} f_n p_n(y) \equiv \sum_{n=0}^{\infty} f_n \phi_n(x), \quad (A2)$$

where $R(y)$ is an entire function of $y$ and $\{f_n\}$ is a new set of expansion coefficients that depend on $\hat{y}$, $\{c_n\}$ and the parameters of the polynomial $p_n(y)$, which is a polynomial of degree $n$ in $y$ with $n$ distinct real zeros. These polynomials in configuration space are not to be confused with the energy polynomials $\{P_n(z)\}$ defined in section II. Therefore, Eq. (A2) allows us to write $\phi_n(x) = R(y-\hat{y})p_n(y)$. Faithful representation of the wavefunction $\psi(E,x)$ dictates that $\{\phi_n(x)\}$ be a complete set and so is $\{p_n(y)\}$. That is,

$$\sum_n \phi_n(x)\phi_n(y) = \frac{\delta(x-y)}{\omega(x)}, \quad (A3)$$

where $\omega(x)$ is a positive definite entire function. This allows us to define the conjugate basis set $\{\bar{\phi}_n(x)\}$ whose elements are

$$\bar{\phi}_n(x) = \omega(x)\phi_n(x) \equiv \bar{R}(y-\hat{y})p_n(y), \quad (A4)$$

where $\bar{R}(y-\hat{y}) = \omega(x)R(y-\hat{y})$. Thus, we can write another series for the conjugate wave function as $\bar{\psi}(E,x) = \sum_{n=0}^{\infty} f_n \bar{\phi}_n(x)$. The orthogonality $\langle\phi_n(x)|\bar{\phi}_m(x)\rangle = \langle\bar{\phi}_n(x)|\phi_m(x)\rangle = \delta_{nm}$ results in the following orthogonality relation for the polynomials $\{p_n(y)\}$

$$\int R(y-\hat{y})\bar{R}(y-\hat{y})p_n(y)p_m(y)\frac{dy}{y'} = \delta_{nm}, \quad (A5)$$

where $y' = dy/dx$. Thus, the weight function $W(y)$ associated with the configuration space polynomials $\{p_n(y)\}$ is

$$W(y) = R(y-\hat{y})\bar{R}(y-\hat{y})/y'. \quad (A6)$$

In most cases, the requirements of square integrability of the basis and the fact that it carries a tridiagonal matrix representation for the wave operator dictate that $[R(y-\hat{y})]^2$ has the same form as the weight function $W(y)$. Now, our choice of the polynomial $p_n(y)$ is limited to those that satisfy the following two criteria:

1. $p_n(y)$ must satisfy a differential equation whose singular points are either $\{y_n\}$ or a subset thereof.
2. The nature and range of the polynomial argument $y$ must be compatible and consistent with that of the physical configuration space coordinate. For example, if $y \geq 0$ then the

–14–

Laguerre polynomial $L_n^v(y)$ whose weight function is $y^v e^{-y}$ becomes a proper choice and we can write $R(y-\hat{y}) = y^\alpha e^{-\beta y}$. Whereas, if $-1 \leq y \leq +1$ then the Jacobi polynomial $P_n^{(\mu,\nu)}(y)$ will be an appropriate choice with the corresponding weight function $(1-y)^\mu (1+y)^\nu$ and we can write $R(y-\hat{y}) = (1-y)^\alpha (1+y)^\beta$. On the other hand, if $-\infty \leq y \leq +\infty$ then the Hermite polynomial $H_n(y)$ with its weight function $e^{-y^2}$ would be a proper choice to take for $p_n(y)$ where we can also write $R(y-\hat{y}) = e^{-\alpha y^2}$.

Based on our past experience, all known exactly solvable potentials correspond to coordinate transformations $y(x)$ such that $y \in [-\infty, +\infty]$, $y \geq 0$, $y \geq a$ or $y \in [a,b]$. These problems are associated with the Hermite polynomial, Laguerre polynomial and Jacobi polynomial. Note that there are many special cases of the Jacobi polynomial like the Legendre, Chebyshev, Gegenbauer, etc.

**APPENDIX B: TRA SOLUTION FOR A NEW CLASS OF POTENTIAL FUNCTIONS**

In this Appendix, we use the second mode of application of the TRA to find the class of all potential functions corresponding to the physical configuration of Section IV and show that the potential well with sinusoidal bottom treated therein is only a special case of a larger class. Therefore, we start by defining the configuration space as the interval $-\frac{L}{2} \leq x \leq +\frac{L}{2}$ and write the one-dimensional Schrödinger equation in terms of the dimensionless coordinate $y = \sin(\lambda x)$, with $\lambda = \pi/L$ and $-1 \leq y \leq +1$, as follows

$$-\frac{\lambda^2}{2}\left[(1-y^2)\frac{d^2}{dy^2} - y\frac{d}{dy} - U(y) + \varepsilon\right]\psi_k(x) = 0, \tag{B1}$$

where $U(y) = 2V(x)/\lambda^2$. The analysis in the Appendix shows that the most general complete basis set appropriate for this problem has the following elements

$$\phi_n(x) = A_n (1-y)^\alpha (1+y)^\beta P_n^{(\mu,\nu)}(y), \tag{B2}$$

where $P_n^{(\mu,\nu)}(y)$ is the Jacobi polynomial and the normalization constant is chosen as $A_n = \sqrt{\frac{2n+\mu+\nu+1}{2^{\mu+\nu+1}} \frac{\Gamma(n+1)\Gamma(n+\mu+\nu+1)}{\Gamma(n+\mu+1)\Gamma(n+\nu+1)}}$. The four dimensionless real parameters are such that $\{\alpha,\beta\}$ positive and $\{\mu,\nu\}$ greater than $-1$. Using the differential equation and differential property of the Jacobi polynomial, we obtain the following matrix representation of the wave operator in this basis [21,31]

$$\frac{2}{\lambda^2}\langle\phi_m|J|\phi_n\rangle = n\langle m|(1-y)^p(1+y)^q\left[p+q-\frac{2p}{1-y}\frac{n+v}{2n+\mu+v}-\frac{2q}{1+y}\frac{n+\mu}{2n+\mu+v}\right]|n\rangle$$

$$-\frac{1}{2}\langle m|(1-y)^p(1+y)^q\left[\frac{(\mu+p)^2-1/4}{1-y}+\frac{(v+q)^2-1/4}{1+y}\right]|n\rangle \quad (B3)$$

$$+\frac{1}{4}\langle m|(1-y)^p(1+y)^q\left[(2n+\mu+v+1)^2+(p+q+1)^2+2(\mu+v)(p+q)-1\right]|n\rangle$$

$$+(2n+\mu+v+1)D_{n-1}\langle m|(1-y)^p(1+y)^q\left(\frac{p}{1-y}-\frac{q}{1+y}\right)|n-1\rangle+\langle m|(1-y)^p(1+y)^q[U(y)-\varepsilon]|n\rangle$$

where $p=2\alpha-\mu-\frac{1}{2}$, $q=2\beta-v-\frac{1}{2}$, $D_n=\frac{2}{2n+\mu+v+2}\sqrt{\frac{(n+1)(n+\mu+1)(n+v+1)(n+\mu+v+1)}{(2n+\mu+v+1)(2n+\mu+v+3)}}$ and we have defined $\langle y|n\rangle = A_n(1-y)^{\mu/2}(1+y)^{v/2}P_n^{(\mu,v)}(y)$. The recursion relation of the Jacobi polynomial and its orthogonality show that this matrix is tridiagonal only in one of three cases:

a) $(p,q)=(0,0)$ and $U(y)=\frac{1}{2}\frac{\mu^2-1/4}{1-y}+\frac{1}{2}\frac{v^2-1/4}{1+y}+U_0 y$. (B4a)

b) $(p,q)=(1,0)$ and $U(y)=\frac{1}{2}\frac{v^2-1/4}{1+y}+\frac{U_-}{1-y}$. (B4b)

c) $(p,q)=(0,1)$ and $U(y)=\frac{1}{2}\frac{\mu^2-1/4}{1-y}+\frac{U_+}{1+y}$. (B4c)

where $U_0$ and $U_\pm$ are arbitrary dimensionless parameters. The relevant case to the present treatment is the first one corresponding to

$$V(x)=\begin{cases}\frac{V_+-V_-\sin(\pi x/L)}{\cos^2(\pi x/L)}+V_0\sin(\pi x/L) &, -\frac{L}{2}\leq x\leq+\frac{L}{2}\\ \infty &, \text{otherwise}\end{cases} \quad (B5)$$

where $V_0$ and $V_\pm$ are real potential parameters, $\mu^2=\frac{1}{4}+\frac{2}{\lambda^2}(V_+-V_-)$, $v^2=\frac{1}{4}+\frac{2}{\lambda^2}(V_++V_-)$ and $U_0=2V_0/\lambda^2$. Reality dictates that $V_+\geq|V_-|-\frac{1}{8}\lambda^2$. It is obvious that the potential well with sinusoidal bottom (21) is a special case with $V_\pm=0$. Note that if $V_0=0$ then the potential function (B5) is just the well-known trigonometric Scarf potential, which is exactly solvable using conventional methods.

Substituting (B4a) in (B3) and using the recursion relation of the Jacobi polynomial and its orthogonality, we obtain an infinite symmetric tridiagonal matrix representing the wave operator with the following elements

$$\frac{2}{\lambda^2}J_{n,m}=\left[\left(n+\frac{\mu+v+1}{2}\right)^2+U_0 C_n-\varepsilon\right]\delta_{n,m}+U_0\left(D_{n-1}\delta_{n,m+1}+D_n\delta_{n,m-1}\right), \quad (B6)$$

where $C_n=\frac{v^2-\mu^2}{(2n+\mu+v)(2n+\mu+v+2)}$. Therefore, the corresponding symmetric three-term recursion relation (9) becomes

–16–

$$\varepsilon P_n^\mu(E) = \left[\left(n + \tfrac{\mu+\nu+1}{2}\right)^2 + U_0 C_n\right] P_n^\mu(E) + U_0 \left[D_{n-1} P_{n-1}^\mu(E) + D_n P_{n+1}^\mu(E)\right]. \quad (B7)$$

Comparing this recursion relation to that of $H_n^{(\mu,\nu)}(z,\sigma)$, which is given as Eq. (E1) of Appendix E in Ref. [31], we conclude that $P_n^\mu(E) = H_n^{(\mu,\nu)}(\varepsilon, U_0)$ and that the bound state wave function (6b) is written as

$$\psi_k(x) = \sqrt{\cos(\lambda x)} \left[1 - \sin(\lambda x)\right]^{\frac{\mu}{2}} \left[1 + \sin(\lambda x)\right]^{\frac{\nu}{2}} \times$$
$$\sqrt{\omega^{(\mu,\nu)}(\varepsilon_k, U_0)} \sum_{n=0}^{\infty} A_n H_n^{(\mu,\nu)}(\varepsilon_k, U_0) P_n^{(\mu,\nu)}(\sin(\lambda x)) \quad (B8)$$

Again, due to the lack of knowledge of the analytic properties of the polynomial $H_n^{(\mu,\nu)}(z,\sigma)$ we resort to numerical means to obtain the physical properties of the quantum system. In Table 3, we give the lowest part of the energy spectrum (in units of $\tfrac{1}{2}\lambda^2$) for a chosen set of values of the potential parameters $V_0$ and $V_\pm$. The Table shows rapid convergence of the calculation with an increase in the basis size.

Finally, if we were to choose either case (B4b) or (B4c), then we would have obtained the class of potentials given by (B5) but with $V_0 = 0$, which is the well-known trigonometric Scarf potential that is exactly solvable using any of the conventional methods. Note, however, that in this case and in contrast to (B5) there is a constraint on the value of only one of the two potential parameters. That is, either $V_+ \geq -(\lambda/4)^2$ or $V_- \geq -(\lambda/4)^2$ while the other is arbitrary and does not have to obey $V_+ \geq |V_-| - \tfrac{1}{8}\lambda^2$. Moreover, the orthogonal energy polynomial in this case becomes the discrete version of the Wilson polynomial whose analytic properties are well known (see, for example, the Appendices in Ref. [30] and citations therein).

**TABLES CAPTION**

**Table 1:** The orthogonal polynomial(s) representing the physical system as a function of the type of its energy spectrum with examples. Note the matching of the spectra of the physical system and that of its corresponding orthogonal polynomial except when the physical system has a mix of a continuous and an infinite discrete energy spectra then it is required to have two polynomials to represent the system: a continuous one and a discrete one with infinite spectrum.

**Table 2:** The lowest part of the energy spectrum (in units of $\frac{1}{2}\lambda^2$) for the infinite potential well with sinusoidal bottom and for the chosen set of values of the potential parameter $\gamma$. We took the basis (23) with 50 elements.

**Table 3:** The lowest part of the energy spectrum (in units of $\frac{1}{2}\lambda^2$) for the potential (B5) with the parameter values $\{V_0, V_+, V_-\} = \{7, 5, 3\}$ in units of $\frac{1}{2}\lambda^2$. Rapid convergence with an increase in the basis size *N* is evident.

**FIGURE CAPTION**

**Fig. 1:** The lowest un-normalized bound-state wavefunctions corresponding to the column $\gamma = 5$ of Table 2. These were calculated using Eq. (28) with the sum converging quickly for the first 7-8 terms. The $x$-axis is in units of $L$ (the width of the well).

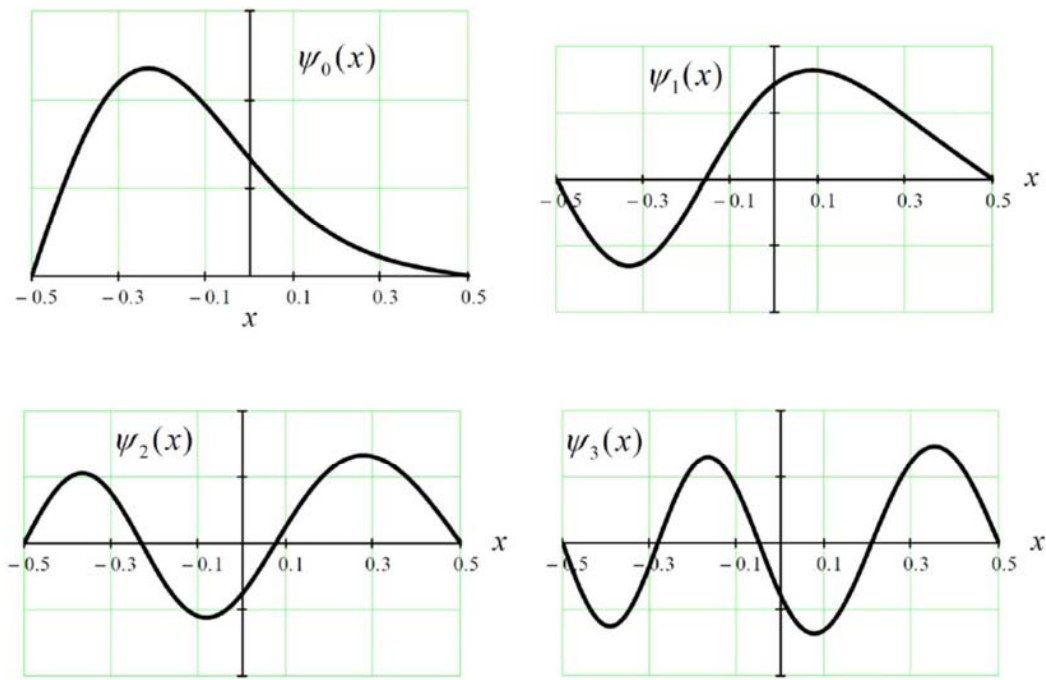

**Fig. 1**

**Table 2**

| n | γ = 0 | γ = 2 | γ = 5 | γ = 10 | γ = 20 |
|---|---|---|---|---|---|
| 0 | 1 | 0.686720257 | −0.595539559 | −3.622765814 | −10.838068721 |
| 1 | 4 | 4.113008823 | 4.345345170 | 3.873494394 | 0.432511407 |
| 2 | 9 | 9.057352856 | 9.354964694 | 10.147416013 | 10.358251307 |
| 3 | 16 | 16.031789784 | 16.200110073 | 16.813060198 | 18.778787010 |
| 4 | 25 | 25.020212925 | 25.126692366 | 25.512098215 | 27.111504117 |
| 5 | 36 | 36.013989568 | 36.087552002 | 36.351914438 | 37.436795310 |
| 6 | 49 | 49.010257797 | 49.064156865 | 49.257285819 | 50.040106169 |
| 7 | 64 | 64.007843753 | 64.049043706 | 64.196465710 | 64.790623174 |
| 8 | 81 | 81.006192252 | 81.038711488 | 81.154988044 | 81.622257081 |
| 9 | 100 | 100.005012691 | 100.0313345580 | 100.125413252 | 100.502864037 |

**Table 3**

| n | N = 10 | N = 11 | N = 12 | N = 13 | N = 100 |
|---|---|---|---|---|---|
| 0 | 7.680625404 | 7.680625404 | 7.680625404 | 7.680625404 | 7.680625404 |
| 1 | 14.338493494 | 14.338493494 | 14.338493494 | 14.338493494 | 14.338493494 |
| 2 | 22.546540967 | 22.546540967 | 22.546540967 | 22.546540967 | 22.546540967 |
| 3 | 32.767801800 | 32.767801800 | 32.767801800 | 32.767801800 | 32.767801800 |
| 4 | 45.034852009 | 45.034852009 | 45.034852009 | 45.034852009 | 45.034852009 |
| 5 | 59.334170173 | 59.334170172 | 59.334170172 | 59.334170172 | 59.334170172 |
| 6 | 75.654554063 | 75.654553948 | 75.654553948 | 75.654553948 | 75.654553948 |
| 7 | 93.988897443 | 93.988866117 | 93.988866057 | 93.988866057 | 93.988866057 |
| 8 | 114.338418785 | 114.332659905 | 114.332639513 | 114.332639480 | 114.332639480 |
| 9 | 137.163172017 | 136.687579697 | 136.683036310 | 136.683022596 | 136.683022577 |

**Table 1**

| Energy Spectrum | | | Potential Example | Polynomial Spectrum | | | Polynomial Example |
|---|---|---|---|---|---|---|---|
| Continuous | Discrete | | | Continuous | Discrete | | |
| | Finite | Infinite | | | Finite | Infinite | |
| ✓ | ✗ | ✗ | Pöschl-Teller: $[\cosh(x)]^{-2}$ | ✓ | ✗ | ✗ | Wilson |
| ✗ | ✗ | ✓ | Oscillator: $r^2$ | ✗ | ✗ | ✓ | Meixner |
| ✓ | ✓ | ✗ | Morse: $e^{-2x} - 2e^{-x}$ | ✓ | ✓ | ✗ | continuous dual Hahn |
| ✓ | ✗ | ✓ | Coulomb: $-r^{-1}$ | ✓ | ✗ | ✗ | Meixner-Pollaczek |
| | | | | ✗ | ✗ | ✓ | Meixner |